\documentclass[aps,twocolumn,superscriptaddress,pra,10pt]{revtex4-1}
\usepackage[OT1]{fontenc}
\usepackage[utf8]{inputenc} 

\usepackage{amsmath}

\allowdisplaybreaks

\usepackage{amssymb}
\usepackage{amsfonts} 
\usepackage{amsmath} 
\usepackage{latexsym} 
\usepackage{array}

\usepackage{graphicx}
\usepackage{wrapfig}  
\usepackage{color}
\usepackage{url}

\usepackage{mathtools}

\RequirePackage{tikz}
\usetikzlibrary{calc}
\usepackage{tikzit}

\definecolor{myred}{HTML}{B50002}
\definecolor{mygreen}{HTML}{81B51E}

\tikzstyle{black}=[fill=black, draw=black, shape=circle]
\tikzstyle{red}=[fill=myred, draw=black, shape=circle]
\tikzstyle{green}=[fill=mygreen, draw=black, shape=circle]
\tikzstyle{redtransp}=[fill=myred, draw=black, shape=circle, opacity=0.5]
\tikzstyle{blacktransp}=[fill=black, draw=black, shape=circle, opacity=0.5]
\tikzstyle{gauge}=[fill=white, draw=black, shape=rectangle, minimum height=0.6 cm]
\tikzstyle{largegaugetransp}=[fill=white, draw=black, shape=rectangle, minimum height=0.6cm, minimum width=1.3 cm, opacity=0.5]
\tikzstyle{greentransp}=[fill=mygreen, draw=black, shape=circle, opacity=0.5]

\tikzstyle{dotted}=[-, dashed]
\tikzstyle{thick}=[-]
\tikzstyle{opaque}=[-, opacity=0.5]

\newcommand{\couic}[1]{}
\newcommand{\couicfootnote}[1]{}
\newcommand{\couicefootnote}[1]{}

\newcommand{\ket}[1]{| #1 \rangle}

\newcommand{\ii}{\mathrm i}

\renewcommand{\ol}[1]{\overline{#1}}

\newcommand{\dyn}{f}
\newcommand{\nol}{}

\newcommand{\bs}{\!\!\!\!\!\!}

\begin{document}

\title{A quantum cellular automaton for one-dimensional QED}

\begin{abstract}
~
\end{abstract}


\author{Pablo Arrighi}
\email{pablo.arrighi@univ-amu.fr}
\affiliation{Aix-Marseille Univ, CNRS, LIS and 
Univ Paris-Saclay, ENS Paris-Saclay, CNRS, INRIA, LSV, France.}
\author{C\'edric B\'eny}
\email{cedric.beny@gmail.com}
\affiliation{Department of Applied Mathematics, Hanyang University (ERICA), 55 Hanyangdaehak-ro, Ansan, Gyeonggi-do, 426-791, Korea.}
\author{Terry Farrelly}
\email{farreltc@tcd.ie}
\affiliation{Institut f\"ur Theoretische Physik, Leibniz Universit\"at Hannover, 30167 Hannover, Germany.}

\begin{abstract}
We propose a discrete spacetime formulation of quantum electrodynamics in one-dimension (a.k.a the Schwinger model) in terms of quantum cellular automata, i.e. translationally invariant circuits of local quantum gates.  These have exact gauge covariance and a maximum speed of information propagation. In this picture, the interacting quantum field theory is defined as a ``convergent'' sequence of quantum cellular automata, parameterized by the spacetime lattice spacing---encompassing the notions of continuum limit and renormalization, and at the same time providing a quantum simulation algorithm for the dynamics.
\end{abstract}

\maketitle

\section{Introduction}

In this work, we propose a discrete spacetime formulation of quantum electrodynamics (QED) in one dimension (the Schwinger model), in terms of quantum cellular automata (QCA), which are essentially translationally invariant circuits of local quantum gates.

From a practical point of view, the QCA defines a quantum simulation algorithm for the dynamics of an interacting QFT (leaving aside the problems of state preparation and measurements, however). But, from a theoretical point of view it also constitutes a proof-of-principle showing that natively discrete formulations of an interacting QFT are possible and elegant. In this picture, the QFT is defined as a ``convergent'' sequence of QCA, parameterized by the spacetime lattice spacing---echoing the notions of continuum limit and renormalization. We discuss why we may hope to circumvent some of the technical issues of standard formulations of QFT this way. 

The construction is intuitive and requires little prerequisites. It leads to a simple, explanatory model of QFT based on quantum information concepts. Given that QFT can be rather intricate, we believe this also constitutes an important pedagogical asset.

\subsection{QFT and their quantum simulation}

Quantum field theory (QFT) is the framework that best describes fundamental particles and their interactions in a relativistic manner \cite{quigg2013gauge}, though without accounting for gravity. QFT encompasses a heterogeneous set of procedures and techniques, but roughly speaking the first step is always to put together a local action, i.e. a way to associate a `cost' to each spacetime history of a classical field. This action is usually derived quite elegantly based on the symmetries of special relativity and the demand that it features local symmetries (e.g., gauge invariance). The result then has to be converted to a quantum theory, through one of several heuristic quantization processes.
Regardless of whether the theory is canonically quantized or quantized via path integrals, handling interactions intrinsically requires a form of regularization.  This is usually equivalent to a discretization in that it involves neglecting small scale features. When one wants to go beyond perturbation theory, recourse to numerical simulations is typically necessary.

Thus, a number of discrete space---or spacetime---formulations of QFT have been studied, understood as approximations designed for numerical simulations. The main simulation method approximately evaluates the path integral in imaginary time (quantum Monte Carlo \cite{HastingsMonteCarlo}). But this only works well in limited cases and, like all classical simulations of quantum theory, suffers from a complexity which is exponential in the number of degrees of freedom.

Confronted with the inefficiency of classical computers for simulating interacting quantum particles, Feynman realized that one ought to use quantum computers instead \cite{FeynmanQC}. What better than a quantum system to simulate another quantum system? In the last decades, several such quantum simulation schemes have been devised \cite{JLP14,georgescu2014quantum, QuantumClassicalSim}, some of which were experimentally implemented recently \cite{InnsbruckLGT}. Generally speaking these are Hamiltonian based \cite{HamiltonianBasedSchwinger}, i.e. they are based on a discrete-space continuous-time version of the QFT, such as the Kogut-Susskind Hamiltonian~\cite{kogut1975hamiltonian}.
One may then look for quantum systems in nature that mimic this Hamiltonian, or that can be tuned into it \cite{ErezCiracLGT}.  Alternatively, one does a staggered trotterization of it so as to obtain unitaries that may be implemented on a digital quantum computer. In any case, by first discretizing space alone and not time, these Hamiltonian-based schemes are essentially taking things back to the non-relativistic quantum mechanical setting:\ Lorentz-covariance is broken. The bounded speed of light can only be approximately recovered (e.g., Lieb-Robinson bounds) \cite{EisertSupersonic,Osborne19}. Even when the Hamiltonian is trotterized, time steps need remain orders of magnitude smaller than space steps. This also creates more subtle problems such as {\em fermion doubling}, where spurious particles are created due to the periodic nature of the momentum space on a lattice. 

An essential aspect of QFT simulations is the preparation of the initial state, boundary conditions, and measurements~\cite{PreskillQuantumSim,JLP14}. This is non-trivial due to the fact that the vacuum of an interacting QFT is not known exactly (or if it is, then we usually also have an exact solution for the evolution). We do not study this question in detail here, except when considering the continuum limit of our QCA in Section~\ref{cont-lim}.

%




\subsection{Quantum simulation via QCA}

From a relativistic point of view, it would be more natural to discretize space and time simultaneously and with the same scale, thereby producing a network of local quantum gates, homogeneously repeated across space and time. Feynman introduced Quantum Cellular Automata (QCA) together with the idea of quantum simulations of physics \cite{FeynmanQCA}. A related model to QCAs was also given by Feynman, namely the simple discretized `checkerboard model' of the electron in discrete $(1+1)$-spacetime \cite{Feynman_chessboard}.

The one--particle sector of QCAs became known as Quantum Walks (QW), and was found to provide quantum simulation schemes for non-interacting fundamental particles \cite{BenziSucci,Bialynicki-Birula,meyer1996quantum,ArrighiDirac}, including in $(3+1)$-spacetime, be it curved \cite{MolfettaDebbasch2014Curved,ArrighiGRDirac,DebbaschWaves,ArrighiGRDirac3D} or not, or in the presence of an electromagnetic field \cite{CGW18}. The sense in which QW are Lorentz-covariant was made explicit \cite{arrighi2014discrete, PaviaLORENTZ, PaviaLORENTZ2, DebbaschLORENTZ}. The bounded speed of light is very natural in circuit-based quantum simulation, as it is directly enforced by the wiring between the local quantum gates. Recently, the two--particle sector of QCA was investigated, with the two walkers interacting via a phase (similar to the Thirring model \cite{DdV87}). This was shown to produce molecular binding between the particles \cite{ahlbrecht2012molecular, PaviaMolecular}.

In the many-particle sector general theorems exist 
showing that simple QCA \cite{ArrighiQGOL} are able to simulate any unitary causal operator \cite{ArrighiUCAUSAL,farrelly2014causal}, and thus arbitrarily complex behaviours. Still, the problem of defining a concrete QCA that would simulate a specific interacting QFT had remained out of reach.  In this paper we give a QCA description of QED in $(1+1)$--spacetime, i.e. the Schwinger model. To the best of our knowledge, this is the first work relating the many-particle sector of a QCA to an interacting QFT. 

\subsection{Towards QCA formulations of QFT}

The apparent simplicity of the standard, path-integral formulation of QFT in fact hides a number of serious complications. These issues are often summarized through the fact that the integration measure over ``paths'' is not mathematically defined, and hence many additional assumptions and techniques have to be used in the process of obtaining actual physical predictions. This implies that the theory is not entirely specified by the action. Other formulations such as canonical quantization (operator formalism) are used in parallel in order to resolve some of these ambiguities. Moreover, whilst Lorentz and gauge invariance are manifest in the path-integral formulation, other fundamental properties are not, i.e. it cannot directly guarantee the unitarity of the dynamics, the existence of a ground state, nor characterize the compatibility of observables.


Another very important source of ambiguities, which cannot be resolved through the canonical operator formalism, comes from the continuity of spacetime. In the classical field theory, this is handled mathematically by the assumption that the fields are adequately differentiable, and formulating the action as an integral of differential quantities. This does not have an equivalent in the quantum setting. Instead, the quantum theory must be first regularized, typically through an energy cutoff, which is essentially equivalent to a discretization of space. Then the continuum limit must be taken on a per-case basis. Different theories, or even different boundary conditions, each require a different highly non-trivial dependence of the parameters on the cutoff (renormalization).

Hence, although the action-based approach has a continuous starting point, it does not actually solve the problem of obtaining a well-defined quantum theory in the continuum:\ the continuum limit needs be taken again in the quantum setting anyway---under the guise of renormalization. Indeed, by itself the action-based approach only succeeds in yielding genuinely continuous quantum theories for (quasi-)free theories. These serve as the basis for perturbative solutions to the renormalization problem---when a weak interaction gets added.

We propose that a natively quantum and discrete formulation of QFT, \`a la QCA, could bypass a number of these issues. 
Since the action-based approach is only apparently continuous anyway,
it may be conceptually clearer to start from a natively discrete theory. (This may also eventually facilitate a connection with genuinely discrete quantum gravity proposals \cite{RovelliLQG,LollCDT}).
Moreover, a genuinely quantum description avoids the ambiguities arising when insisting on developing a quantum theory around a classical action.  QCA have some other desirable properties:\ they are manifestly unitary and they also come with the immediate advantage of having a local formulation, from which an strict limit on the speed of information propagation can be extracted, as in special relativity. Lorentz-covariance can be handled as in \cite{ArrighiLORENTZ, PaviaLORENTZ, PaviaLORENTZ2, DebbaschLORENTZ}. Gauge-invariance can also be handled as in \cite{ArrighigaugeRCA}, a technique which is inspired by 
gauge-invariant quantum walks \cite{di2014quantum,arnault2016quantum,di2016quantum}. This technique allows one to construct the interacting theory from the free theory and the gauge symmetry requirement, reminiscent of the way the Hamiltonian of a lattice gauge theory gets constructed \cite{rico2014tensor,silvi2014}.

The present paper illustrates precisely this point. Our point of departure is the simple and well-known Dirac QCA for free fermions. Based on the demand that it features a local $U(1)$--gauge symmetry, a gauge field gets introduced that `counts' fermions. Interactions then arise by having fermions pick up a local phase when they move left or right.
Renormalization comes into play in the way the QCA parameters are made to depend on the lattice spacing, leading to renormalization trajectories.





\subsection{The Schwinger model}


The Schwinger model, i.e.\ $(1+1)$--QED, is a useful playground to understand phenomena in QFT in higher dimensional spaces. Let us give a rapid summary of its main features, based on Ref.~\cite{melnikov2000lattice}. This subsection could be skipped by readers that are only interested in the QCA itself, i.e. not interested in comparing it with standard $(1+1)$--QED.

The Schwinger model can be formulated covariantly in terms of an action, but let us directly write down a Hamiltonian formulation, which will be easier to compare to our QCA. The Hamiltonian formalism requires the local gauge symmetry to be partially fixed, which can be done in many ways. In the temporal gauge (with $A_0(x)=0$, and writing $A(x)=A_1(x)$), the Hamiltonian is $H=$
\begin{equation}
\label{SchwingerHam}
    \int \!\mathrm{d}x \bigl(\psi^{\dagger}(x)\left[\left(i\partial_x+i e A(x)\right)\sigma_z+m\sigma_x\right]\psi(x)+\frac{1}{2}E^2(x)\bigr),
\end{equation}
where $E(x)$ is the electric field observable at $x$, and $A(x)$ is its conjugate momentum, meaning
\begin{equation}
    [A(x),E(y)]=i\delta(x-y).
\end{equation}
Here $\psi(x)=(\psi_1(x),\psi_2(x))^T$ is a two component fermion field satisfying
\begin{equation}
    \{\psi_{\alpha}(x),\psi^{\dagger}_{\beta}(y)\}=\delta_{\alpha\beta}\delta(x-y),
\end{equation}
The electric charge is denoted by $e$, and the mass of the fermions is $m$.

{Choosing the temporal gauge does not completely fix the gauge freedom. One still needs to prevent that the remaining gauge degrees of freedom be observed. One way of doing this is to demand that physical states commute with Gauge transformations, making them ``block-diagonal''. In terms of pure states, this amounts to forbidding superposition across any two distinct eigenspaces of a Gauge transformation. In the present context this leads to Gauss' law:
\begin{equation}
    \left[\partial_x E(x)-e\psi^{\dagger}(x)\psi(x)\right]\ket{\phi_{\mathrm{phys}}} = f(x)\ket{\phi_{\mathrm{phys}}}.
\end{equation}
where the $f(x)$ is fixed. Mathematically it corresponds to the choice of an eigenspace, physically it can be interpreted as an external, fixed electric field. Notice that because the operators implementing Gauss' law commute with $H$, initial states satisfying Gauss' law also satisfy it in the future. An alternative is to restrict the set of observables instead, and demand that these commute with Gauge transformations.}

The Schwinger model is phenomenologically very different from QED in three dimensions, but it exhibits many interesting phenomena that arise in other QFT in $(3+1)$--dimensions.  The phenomena appearing depend on whether the fermion field is massive or massless.  (Some authors reserve the name Schwinger model for the massless case.)   For example, massless fermions suffer confinement:\ the effective electric force between particles of opposite charges increases with distance, consequently charges are paired into an effective particle: a massive boson, whose mass is proportional to the electromagnetic coupling strength. 


\subsection{Plan}

In order to construct the Schwinger QCA we replay the procedure for constructing QED, in a natively discrete manner. In Section \ref{sec:dirac} we start with the one-particle sector, namely the Dirac QW, which simulates the Dirac Eq.\ for a free electron in a simple and elegant manner, only upgraded to the many non-interacting particle sector, yielding a Dirac QCA. In Section \ref{sec:gaugeinv} we extend the Dirac QCA minimally, so that it acquires a local $U(1)$ symmetry, thereby introducing the electromagnetic field---this cellular automata version of the gauging procedure was developed in \cite{ArrighigaugeRCA}. In Section \ref{sec:interaction} the QCA is endowed with a simple local phase gate, which implements the interaction. Throughout the paper, we consider both the Schr\"odinger and Heisenberg pictures (discussing fermionic creation and annihilation operators, how they evolve, and which are the physical observables). Finally we sketch the continuum limit. Section \ref{sec:conclusion} summarizes the results and puts them into perspective.


\section{Dirac QCA}\label{sec:dirac}

\begin{figure}
\includegraphics[width=0.8\columnwidth]{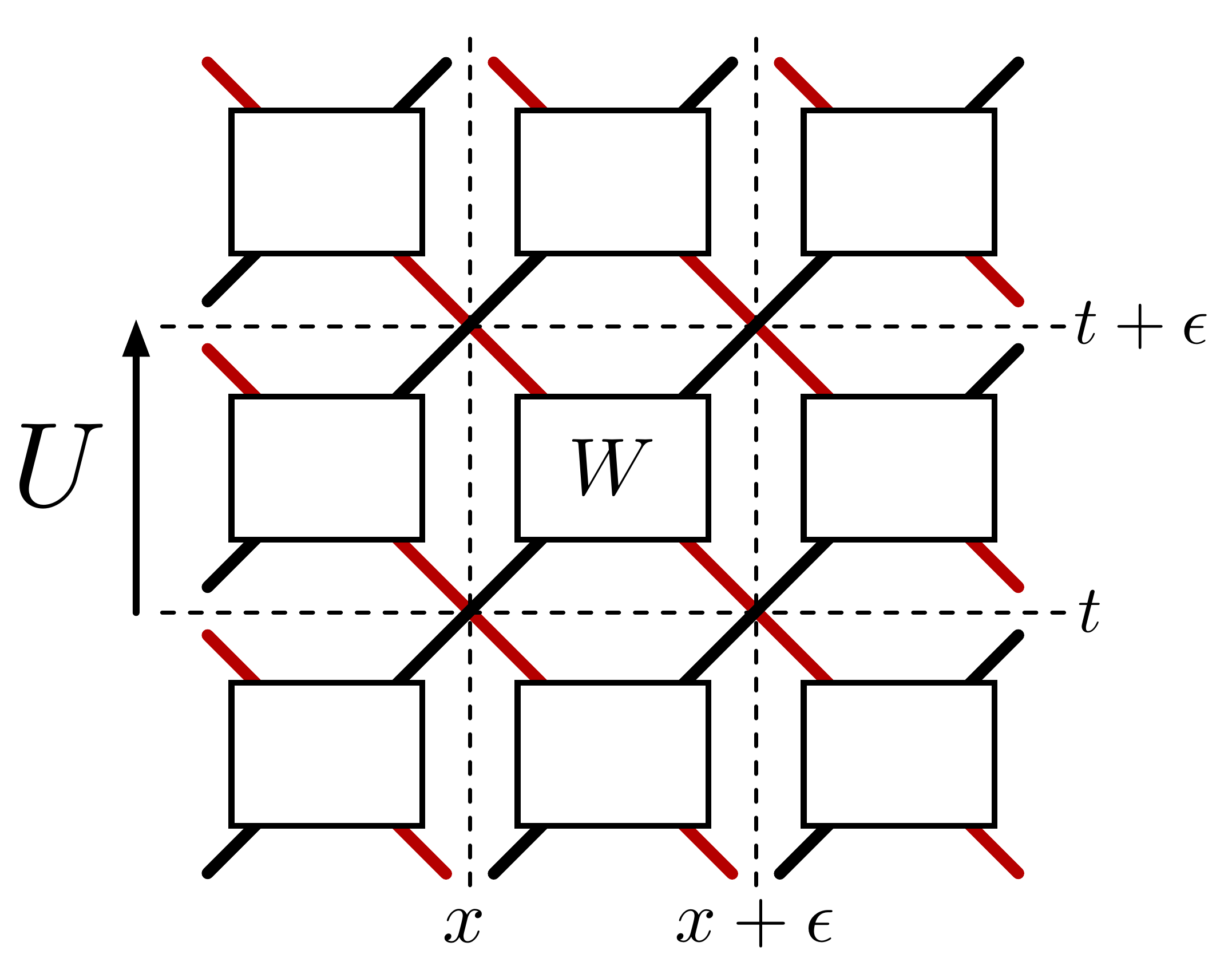}
\caption{\label{fig:DiracQCA}
Dirac QCA: each space-time point, the intersections of the dotted lines, $(x,t)$ can be occupied by a left and a right-moving fermion. All gates are identical and given by the matrix $W$, which allows the fermions to change direction with an amplitude that depends on its mass. Because this is a qubit description, each crossing of two fermions triggers a $-1$ phase, including within the gate $W$.}
\label{figex}
\end{figure}

\subsection{Schr\"odinger picture and qubit representation}

We arrange the Dirac QCA as shown in Fig.~\ref{fig:DiracQCA}. We have two Dirac fermions at each site $x = \varepsilon k$, $k \in \mathbb Z$ of a one-dimensional lattice (where the red and black wires cross), which can be thought of as having orthogonal ``chirality'', i.e. that are about to move in opposite directions.  There are two ways to describe the QCA:\ via fermions or via qubits (related by the Jordan-Wigner isomorphism).  First, we will give the qubit description, and later we will give the QCA in the fermionic picture.

We encode the occupation number of each fermion as a qubit, yielding two qubits at each site. The red (resp.\ black) wires denote the qubit corresponding to the occupation number of the left-moving (resp.\ right-moving) modes.

All gate are identical, with components
\begin{align}
&W= \left(
\begin{array}{cccc}
1 & 0 & 0 & 0\\
0 & -\ii s & c & 0\\
0 & c & -\ii s & 0\\
0 & 0 & 0 & -1
\end{array}
\right)\\
&=1\oplus \sigma_1 \exp(-\ii m\varepsilon \sigma_1) \oplus -1\label{eq:DiracGate}
\end{align}
with $c=\cos(m \varepsilon)$ and $s=\sin(m \varepsilon)$ accounting for the mass.   

The components are ordered so that when the mass is zero, the particles do not change direction, so that a right-moving mode is transferred from $x$ to $x+\varepsilon$, etc.

The minus sign of the bottom-right entry of the matrix is needed in the qubit representation of two fermions crossing past each other, as we shall see. For the same reason, each crossing of a red and black wire also has a gate
\begin{equation}
S = \begin{pmatrix} 
1 & 0 & 0 & 0\\ 0 & 1 & 0 & 0 \\ 0 & 0 & 1 & 0 \\ 0 & 0 & 0 & -1
\end{pmatrix}.
\end{equation}

\subsection{Single particle dynamics}\label{subsec:oneparticledyn}

To see that this QCA is a discretization of non-interacting Dirac particles, let us consider the ``one-particle'' sector, by restricting the QCA to the subspace spanned by states with a single fermionic mode occupied.  In other words, we consider states with all but one qubit in the zero state.  Define $\ket{x+}$ to be the state with the qubit corresponding to a right mover at position $x$ being in the $1$ state and zeroes everywhere else.  Similarly $\ket{x-}$ is the state with the left-mover qubit at position $x$ being in the $1$ state with zeroes everywhere else.  Then we consider the state at time $t$ given by
\begin{equation}
    \ket{\psi(t)}=\sum_x\left[\psi^+(t,x)\ket{x+}+\psi^-(t,x)\ket{x-}\right],
\end{equation}
with $\ket{\psi(t)}$ normalized to one.  Then the unitary dynamics gives us the update rule after one timestep to be
\begin{equation}\label{eq:sp}
\begin{split}
\psi^+(t+\varepsilon,x)&=c\psi^+(t,x-\varepsilon)+ -\ii s\psi^-(t,x)\\ 
\psi^-(t+\varepsilon,x)&=c\psi^-(t,x+\varepsilon)+ -\ii s\psi^+(t,x).
\end{split}
\end{equation}
A rough justification of the continuum limit, is simply to expand to first order in $\varepsilon$, giving
\begin{align}
\varepsilon\partial_t \psi^+&=-\varepsilon\partial_x \psi^+-\ii m\varepsilon\psi^-\\ 
\varepsilon\partial_t \psi^-&=+\varepsilon\partial_x \psi^--\ii m\varepsilon\psi^+.
\end{align}
Dividing across by $\varepsilon$ allows us to write this in terms of Pauli matrices to get
\begin{align}
\partial_t \psi&=-\sigma_3\partial_x\psi -\ii m\sigma_1\psi,\label{eq:Dirac}
\end{align}
which is the Dirac equation, where $\psi=(\psi^+,\psi^-)^T$.  More rigorous treatments of the continuum limits of such single-particle dynamics (i.e., quantum walks) can be found in, e.g., \cite{arrighi2013dirac}.

\subsection{Heisenberg picture and fermion representation}

Let us look at the QCA in terms of fermion operators, which will allow us to show that the evolution is consistent with free Dirac QFT.

The quasi-local algebra of operators for free Dirac QFT is generated by the finite products of the operators $a_x$, $b_x$, $a_x^\dagger$ and $b_x^\dagger$ for all $x$, with $a_x$ (resp. $b_x$) the annihilation operator of a left-moving (resp. right-moving) mode at each point $x$. Since the adjoint evolution of the free Dirac QFT is an automorphism of the $*$-algebra, it is entirely characterized by the way it acts upon $a_x$, $b_x$, over an $\varepsilon$--period of time. It naturally arises as the second quantization of Subsection \ref{subsec:oneparticledyn}:
\begin{equation}
\label{eq:dirac-heisenberg}
\begin{split}
\dyn(a_x^\dagger) &= \cos(m\varepsilon)\, a_{x+\varepsilon}^\dagger + \ii \sin(m\varepsilon) \, b_x^\dagger \\
\dyn(b_x^\dagger) &=  \ii \sin(m\varepsilon) \, a_{x}^\dagger + \cos(m\varepsilon) \, b_{x-\varepsilon}^\dagger. \\
\end{split}
\end{equation}  
This could have been the basis for the definition of a fermionic QCA \cite{PaviaMolecular} in the Heisenberg picture \cite{SchumacherWerner}. 
The relation to the qubit QCA in the previous section can be established as follows. 
First notice that the $f$ defined by Eqs.~\eqref{eq:dirac-heisenberg} is occupation-number-conserving.
Now say we had an occupation-number-conserving unitary $U$ such that $\dyn(A) = U^\dagger A U$. We would then have 
\begin{equation}
\dyn(A)\ket 0 = U^\dagger A U \ket 0 = U^\dagger A \ket 0,\label{eq:qcaspecs}
\end{equation} 
where $\ket{0}$ denotes the vacuum for the modes $a_x$ and $b_x$. This equation would then specify $U$ from $f$. Since $f$ is local, we can directly deduce $W$ instead, again assuming it is occupation-number-conserving. Indeed, let us write the four possible input states of a gate $W$ as
\begin{equation}
\ket{m,n} = (b_x^\dagger)^m (a_{x+\varepsilon}^\dagger)^n \ket{0}.
\end{equation}
with $m$ being the number of right-moving fermions at $x$ (zero or one, created by $b_x^\dagger$) and $n$ the number of left-moving fermions at $x+\varepsilon$ (zero or one, created by $a_{x+\varepsilon}^\dagger$). 
Similarly, we write the four output states as 
\begin{equation}
\ket{n,m}' = (a_x^\dagger)^n (b_{x+\varepsilon}^\dagger)^m \ket{0}.
\end{equation}
Occupation number conservation establishes the first column of the matrix $W^\dagger$. For the second column, the output state $\ket{01}' = b_{x+\varepsilon}^\dagger \ket{00}'$ must be mapped by $W^\dagger$ to 
\begin{equation}
\dyn(b_{x+\varepsilon}^\dagger)\ket{00} 
= (\ii s \,a_{x+\varepsilon}^\dagger + c\, b_{x}^\dagger)\ket{00} 
= \ii s \,\ket{01} + c \,\ket{10},
\end{equation}
and similarly for the third column.
For the last column, the state $\ket{11}'$ must be mapped by $W^\dagger$ to 
\begin{equation}
\begin{split}
\dyn(a_x^\dagger b_{x+\varepsilon}^\dagger)\ket{00}
&= (c \,a_{x+\varepsilon}^\dagger + \ii s\, b_x^\dagger ) (\ii s \,a_{x+\varepsilon}^\dagger + c \,b_x^\dagger )\, \ket{00}\\
&= (c^2 a_{x+\varepsilon}^\dagger b_x^\dagger - s^2 b_x^\dagger a_{x+\varepsilon}^\dagger)\, \ket{00}\\
&= - (c^2 + s^2)\, b_x^\dagger a_{x+\varepsilon}^\dagger \ket{00}\\
&= - \ket{11},
\end{split}
\end{equation}
which justifies the $(-1)$ of Eq. \eqref{eq:DiracGate}.


\subsection{Fermion doubling}\label{sec:doubling}

\begin{figure}
	\includegraphics[width=0.8\columnwidth]{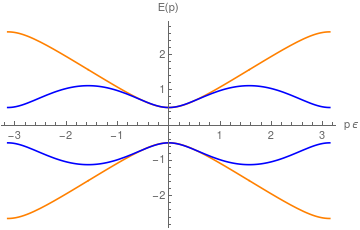}
	\footnotesize{\caption{Dispersion relations for the fermion QCA (orange) and for naive fermions in lattice QFT (blue).  For small momentum, both dispersion relations look like those for Dirac fermions $E(p) = \pm \sqrt{p^2+m^2}$.  Notice that the naive fermions also have low energy modes around $p\varepsilon=\pi$.  These correspond to doubler modes.}\label{fig:spectra}}
\end{figure}

This subsection is just to discuss well-known technical difficulty that sometimes arises when discretizing relativistic fermions, called the fermion doubling problem.  This occurs when the discrete model has more low energy modes than one wants in the continuum \cite{DGDT06}. An illustrative example of this is so-called naive fermions in lattice quantum field theory.  These are described by the Hamiltonian
\begin{equation}
 H = \sum_{k=0}^{N-1}\psi_k^{\dagger}\left(\frac{\sin(k)}{\varepsilon}\sigma_z+m\sigma_x\right)\psi_k,
\end{equation}
where $\psi_k=(a_{k},b_{k})$ are the fermion annihilation operators in momentum space and $k$ labels lattice momenta.  The dispersion relation is given by
\begin{equation}
 E(p)=\pm\sqrt{\sin(k)^2/\varepsilon^2+m^2},
\end{equation}
where $p=k/\varepsilon$ corresponds to the continuum momentum in the continuum limit.  This is plotted in Fig. \ref{fig:spectra} and we see that, for small $k=p\varepsilon$, $E(p)$ is close to the relativistic dispersion relation $E(p) = \pm \sqrt{p^2+m^2}$.  However, looking at momenta $p\varepsilon=\pi+\delta$, the dispersion relation also looks like $\pm\sqrt{\delta^2+m^2}$.  So there are extra particles that behave like relativistic particles.

This is not a problem for free theories.  If the initial state has only low momentum modes occupied, then this will not change as the system evolves.  In contrast, when there are interactions, then high momentum, low energy particles may be created, even if the initial state has only low momentum modes occupied, which would affect, e.g., scattering amplitudes.

That was an example of fermion doubling for massive fermions.  In the massless case for local Hamiltonians on lattices, this is unavoidable as a consequence of the Nielsen-Ninomiya theorem \cite{NN81,DGDT06}, without losing local chiral symmetry, or using other  complicated tricks, e.g., domain wall fermions \cite{Kaplan09}.  And the naive fermion example here illustrates that the problem can even arise for some discretizations of massive fermions.  Let us consider whether this occurs for our model.

Looking at equations (\ref{eq:sp}), we can write the evolution in the single particle picture as
\begin{equation}
 U=\begin{pmatrix}
    \cos(m\varepsilon)S & -i\sin(m\varepsilon) \\
    -i\sin(m\varepsilon) & \cos(m\varepsilon)S^{\dagger}
   \end{pmatrix},
\end{equation}
where $S$ shifts the particle by one lattice step.  In momentum space this is $S=e^{-ik}=e^{-ip\varepsilon}$, so we can find the eigenvalues of $U$ for each value of momentum:
\begin{equation}
 \Lambda_{\pm}(p)=\cos(m\varepsilon)\cos(p\varepsilon)\pm\sqrt{\cos(m\varepsilon)^2\cos(p\varepsilon)^2-1}.
\end{equation}
If we define the quasi-energies by $E(p)=i\ln[\Lambda_{\pm}(p)]/\varepsilon$, then we see from the plot in Fig. \ref{fig:spectra}, that fermion doubling does not occur in the spectra for these models.

Other works have considered similar discrete-time models in the free case \cite{DdV87,Farrelly15,FML17}, where they reached the same conclusion, namely that these models do not suffer from the fermion doubling problem.  However, a more detailed analysis in the presence of interactions would be extremely useful.

\section{Imposing gauge-invariance}\label{sec:gaugeinv}

\begin{figure}
\includegraphics[width=0.8\columnwidth]{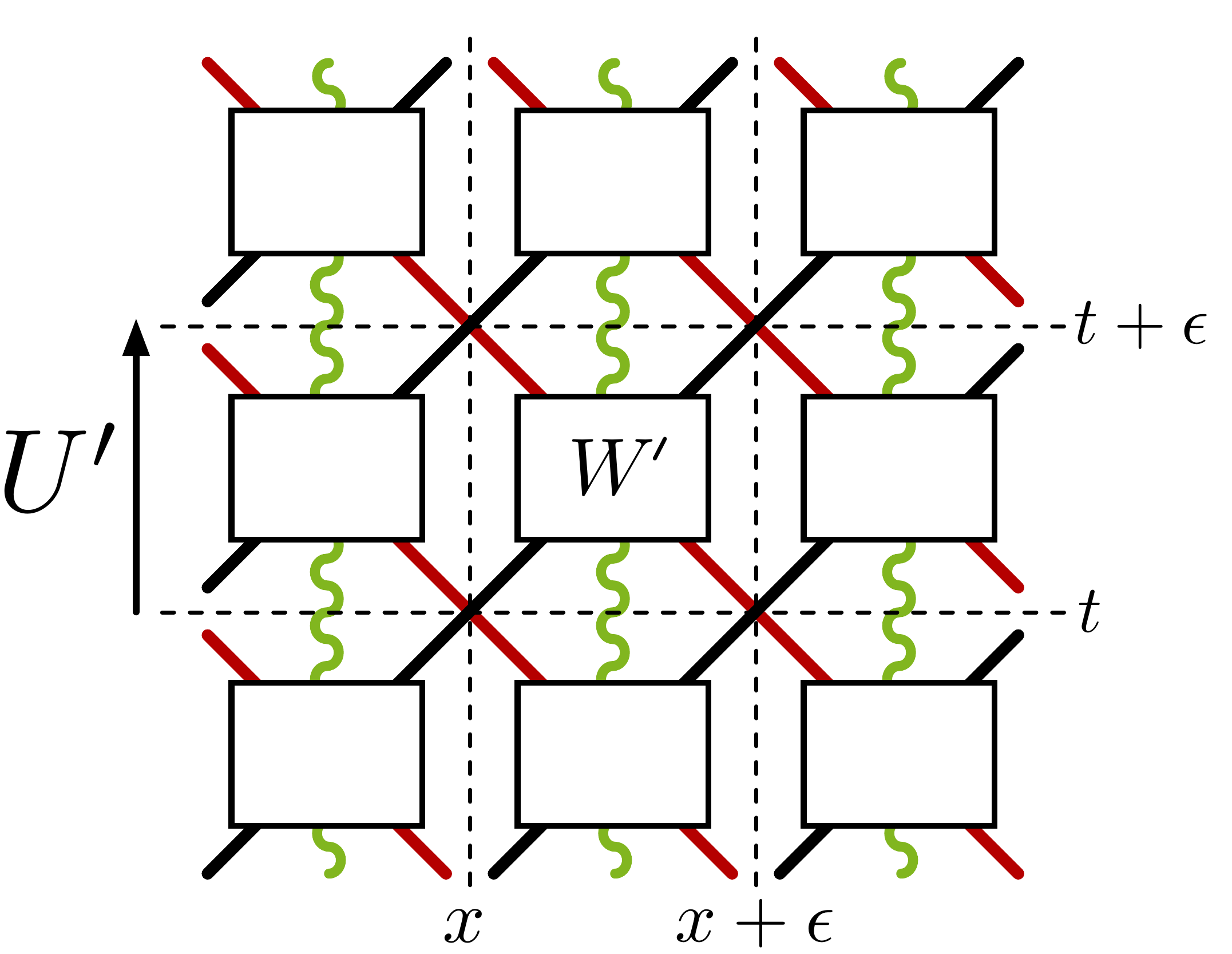}
\caption{\label{fig:QEDQCA} $(1+1)$--QED QCA structure. At $x+\varepsilon/2$ positions lies a wire carrying a state in ${\cal H}_{\mathbb{Z}}$ representing the gauge field. Its sole role is to count the fermions passing by, and to undergo a phase accordingly:\ this phase triggers the interaction. }
\label{figexg}
\end{figure}

\subsection{Schr\"odinger picture}

\begin{figure}
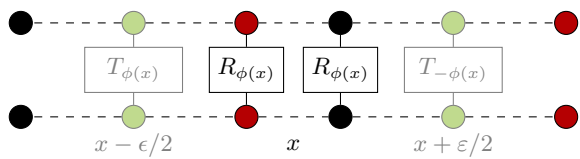

\ctikzfig{Gphi}
\caption{\label{fig:Gphi} The gauge transformation.}
\end{figure}

We want to enforce a local phase symmetry. The symmetry group's elements are specified by a field $\varphi:\mathbb{\varepsilon \mathbb Z}\rightarrow \mathbb{R}$ stating which phase to apply at each point in space, see Fig. \ref{fig:Gphi} (black boxes). The induced, {\em gauge transformation} acts on a state $\psi$ according to
\begin{align*}
G_\varphi \psi&= \big(\bigotimes_x G_\varphi(x)\big)\psi \quad\textrm{with}\quad G_\varphi(x)=R_{\varphi(x)}\otimes R_{\varphi(x)}
\end{align*}
\begin{align*}
R_\phi : \ket{0} &\mapsto \ket{0}\nonumber\\
\ket{1} &\mapsto e^{i\phi}\ket{1}\nonumber
\end{align*}
In other words, both the red and black qubits at $x$, corresponding to the left and right-moving fermions occupation numbers at this discrete position, get multiplied by the phase $\phi(x)$ when they are in the $\ket{1}$ state.

This is the symmetry we are trying to enforce. We want to extend the Dirac QCA $U$ into a {\em gauge-invariant} QCA $U'$, such that for any $\varphi$, there exists a $\varphi'$ such that 
\begin{align}
G_{\varphi'} U'=U' G_{\varphi} .\label{eq:gaugeinv}
\end{align}
In order to turn this gauge-invariance condition into a local one, we use the fact that $U$ has the quantum circuit structure of Fig. \ref{fig:DiracQCA}. This simplification is not specific to our case---every QCA can be made to have that structure, see \cite{ArrighiPQCA}. The obtained local condition is that for all $\varphi$, there exists $\varphi'$, such that for all $x$, 
\begin{align}
\big(R_{\varphi'(x)}\otimes R_{\varphi'(x+\varepsilon)} \big) W' = W' \big(R_{\varphi(x)} \otimes R_{\varphi(x+\varepsilon)}\big).\label{eq:localgaugeinv}
\end{align}

This is not the case for the Dirac QCA as it stands, even when $m=0$. Indeed say our input is $\ket{10}+\ket{01}$, we are asking that there be a $\varphi'$ such that $e^{\ii\varphi'(x)}\ket{10}+e^{\ii\varphi'(x+\varepsilon)}\ket{01}$ be equal to 
\begin{align}
&W\big( e^{\ii\varphi(x)}\ket{10} + e^{\ii\varphi(x+\varepsilon)}\ket{01} \big)\\
&= e^{\ii\varphi(x+\varepsilon)}\ket{10}+e^{\ii\varphi(x)}\ket{01},
\end{align}
thereby imposing on $\varphi'$ the contradictory requirements that for all $\varphi$ and $x$, $\varphi'(x)=\varphi(x+\varepsilon)$ and $\varphi'(x+\varepsilon)=\varphi(x)$.

In order to fix this, we follow the prescription of lattice gauge theory as follows: we interleave ancillary cells on the edges of the line graph, i.e.\ at positions $x+\varepsilon/2$. These cells carry an integer $l\in\mathbb{Z}$, its Hilbert space is ${\cal H}_{\mathbb{Z}}$. This is the so-called {\em gauge field}. Let us extend the gauge transformation to become 
\begin{align}
G_\varphi(x)= T_{\varphi(x)}\otimes R_{\varphi(x)}\otimes R_{\varphi(x)}\otimes T_{-\varphi(x)}
\end{align}
with $T_{\varphi(x)}\ket{l}=e^{\ii l\varphi(x)} \ket{l}$ acting on the left and right edges at $x\pm \varepsilon/2$, see Fig. \ref{fig:Gphi} (now including gray boxes). These $G_\varphi(x)$ no longer have disjoint supports, but they do commute, because $T_{-\varphi(x)}$ commutes with $T_{\varphi(x+\varepsilon)}$. Thus $G_\phi = \prod_x G_{\phi(x)}$ is well-defined. At $x+\varepsilon/2$, it ends up mapping $\ket{l}$ into $\ket{l}=e^{\ii l(\varphi(x+\varepsilon)-\varphi(x))}\ket{l}$. Let us extend $W$ into a $W'$ that affects the gauge field, namely
\begin{align}
&W'= \left(
\begin{array}{cccc}
{I} & 0 & 0 & 0\\
0 & -\ii s {I} & c {V} & 0\\
0 & c V^\dagger & -\ii s {I}& 0\\
0 & 0 & 0 & -{I}
\end{array}
\right)\label{eq:IntermGate}
\end{align}
with ${V}\ket{l}=\ket{l-1}$ and ${I}$ the identity.\\
This, in turns out, is gauge-invariant with just $\varphi'=\varphi$. In order to prove it let us focus on the input $\ket{mln}$ of a gate. When applying $G_\varphi$, this input state will trigger a phase gain $\vartheta(x,m-l,n+l)$: 
\begin{align}
&m\varphi(x)+l(\varphi(x+\varepsilon)-\varphi(x))+n\varphi(x+\varepsilon)\\
&=(m-l)\varphi(x)+(n+l)\varphi(x+\varepsilon).
\end{align}
Now, observe that the numbers $(m-l,n+l)$ are invariants of $W'$, as it takes $\ket{mln}$ into a superposition of the form
\begin{equation}
\sum_{i\in\{-1,0,1\}} \alpha_i \ket{m-i,l-i,n+i}.
\end{equation}
It follows that, when applying $G_{\varphi'}=G_{\varphi}$, the output state will trigger the same phase gain $\vartheta(x,m-l,n+l)$.
Thus, we restored gauge-invariance at the price of introducing ancillary cells in ${\cal H}_{\mathbb{Z}}$. We may wonder whether finite-dimensional alternatives exist, that would be more in line with the QCA tradition. One argument for this will given in Subsec. \ref{subsec:schrodinteraction}.

\subsection{Gauge-invariant observables}

In a gauge theory, only those observables which are invariant under all gauge transformations are physical. Let us characterize these observables. This is not strictly necessary for constructing the QCA, but it is instructive to see how we could actually define the QCA in the Heisenberg picture in a purely gauge-invariant manner, via an automorphism of the algebra of physical observables.

Let 
\(
A_x = a_x^\dagger a_x 
\)
and
\(
B_x = b_x^\dagger b_x 
\)
be the number operators for the two fermionic modes at the point $x$. 

Remember that the gauge field lies at positions $x \in \varepsilon \mathbb Z + \frac 1 2 \varepsilon$. On the edge $x$ between vertex $x-\varepsilon/2$ and $x+\varepsilon/2$, we define the operator $L_{x}$ by $L_{x} \ket l = l \ket l$. (The ``electric field'' at site $x$ is $E_x = g L_x$, where $g$ is the charge of our particles).
Together with $V_{x}$, it generates the algebra of operators for the gauge field on that edge, which is defined algebraically by $[V_x, L_x] = V_x$.

The operators 
\begin{equation}
J_x := L_{x+\varepsilon/2} - L_{x-\varepsilon/2} - A_x - B_x
\end{equation}
for all $x$ generate the group of gauge transformations, at a given time. Specifically,
\begin{equation}
G_\varphi = e^{-i \sum_x \varphi(x) J_x}.
\end{equation}

As in Ref.~\cite{melnikov2000lattice}, let us consider the operators
\begin{align}
\ol a_x &= a_x \Pi_{y > x} V_y\\
\ol b_x &= b_x \Pi_{y > x} V_y\\
\end{align}
whose adjoints create fermions while increasing the electric field to their right. These operators are not local, but we will use them to build physical local observables. 

Since the operators $V_x$ commute with each other and with $a_x$ and $b_x$, then $\ol a_x$ and $\ol b_x$ satisfy the same anticommutation relations as $a_x$ and $b_x$, namely,
\begin{equation*}
\begin{split}
\{\ol a_x, \ol a_y^\dagger\} &= \{\ol b_x, \ol b_y^\dagger\}=\delta_{x,y}I\\
\{\ol a_x,\ol a_y\} &= \{\ol b_x,\ol b_y\} = \{\ol a_x,\ol b_y\}= \{\ol a_x,\ol b_y^\dagger\} = \{\ol a_x^\dagger,\ol b_y\} = 0  
\end{split}    
\end{equation*}

Moreover,
\begin{equation}
\label{cr}
[\alpha \ol a_x + \beta \ol b_x, L_y] = 
\begin{cases}
\alpha \ol a_x + \beta \ol b_x & \text{if $x < y$}\\
0 & \text{otherwise}
\end{cases}
\end{equation}
for any $\alpha, \beta \in \mathbb C$.

The point of these new fermionic annihilation operators is that they commute with all the generators of gauge transformations $J_x$. Together with the electric field operators $L_x$, they generate the set of operators commuting with the gauge generators $J_x$ for all $x$.

However, we could argue that a much smaller algebra is physical, if we only allow local operators, and only even order polynomials in fermionic creation or annihilation operators. This suggests restricting observables to the quasilocal $C^*$-algebra $\mathfrak A$ generated by the operators $L_x$ together with
\begin{align}
\ol a_x^\dagger \ol a_x &= a_x^\dagger a_x\label{eq:localg1}\\
\ol b_x^\dagger \ol b_x &= b_x^\dagger b_x\\
\ol a_x^\dagger \ol a_{x+\varepsilon} &= a_x^\dagger V_{x+\frac 1 2 \varepsilon}^\dagger a_{x+\varepsilon}\\
\ol b_x^\dagger \ol b_{x+\varepsilon} &= b_x^\dagger V_{x+ \frac 1 2 \varepsilon}^\dagger b_{x+\varepsilon} \\
\ol b_x^\dagger \ol a_x &= b_x^\dagger a_x\label{eq:localg5}
\end{align}
for all $x$. 
Indeed, observe that all even polynomials are generated by order two polynomials. Moreover, for such terms to be local, they must create and annihilate one fermion. Products of fermionic operators further apart can be obtained by anticommuting the above terms. For instance:
\begin{equation}
\{\ol a_{x}^\dagger \ol a_{x+\varepsilon}\, ,\, \ol a^\dagger_{x+\varepsilon} \ol a_{x+2\varepsilon}\} = \ol a_x^\dagger \ol a_{x+2\varepsilon}.
\end{equation}

\subsection{Heisenberg dynamics}

A QCA defined by a map $\dyn$ is gauge invariant, as defined in the previous section, if for all $\varphi$ there exists $\varphi'$ such that, for all observable $X$,
\begin{equation}
\dyn(G_{\varphi'}^\dagger X G_{\varphi'}) = G_{\varphi}^\dagger \dyn(X) G_{\varphi}.
\end{equation}
For $X$ gauge-invariant, this reduces to
\begin{equation}
\dyn(X) = G_{\varphi}^\dagger \, \dyn(X) \, G_{\varphi}
\end{equation}
for all $\varphi$, i.e., $\dyn(X)$ must be gauge-invariant too.
Hence, this implies that $\dyn$ must be an automorphism of the gauge-invariant algebra.


We can obtain such an automorphism $\dyn$ by simply substituting the new fermionic operators in Eq.~\eqref{eq:dirac-heisenberg}:
\begin{equation}
\label{eq:eoma}
\begin{split}
\dyn(\ol a_x) &= c \,\overline a_{x+\varepsilon} - i s \,\overline b_x\\
\dyn( \ol b_x) &= - i s \,\overline a_x + c \,\overline b_{x - \varepsilon}
\end{split}
\end{equation}
where $c = \cos(\varepsilon m)$ and $s = \sin(\varepsilon m)$. 

But we still have to chose how $f$ acts on $L_x$. The QCA defined in the previous section, for instance, is compatible with Eq.\eqref{eq:eoma}, and gives us
\begin{equation}
\label{eq:eomb}
f(L_x) = L_x + A_{x+ \frac 1 2 \varepsilon } - f(B_{x+\frac 1 2 \varepsilon }).
\end{equation}

This has the property that it fixes the gauge generators:
\begin{equation}
f(J_x) = J_x \quad \text{for all $x$},
\end{equation}
which is how the gauge choice $\varphi_x' = \varphi_x$ manifests in this picture. To check this equation, it is useful to use the fact that the number of fermions is locally conserved in the sense that
\begin{equation}
\dyn(A_x)+\dyn(B_{x+\varepsilon}) = A_{x+\varepsilon} + B_x.
\end{equation}

One could verify that this map $f$ is local by showing that the image of the local generators are local. However this follows also from the facts that it is implemented by the finite circuit of Fig.~\ref{fig:QEDQCA}.

\section{Interacting QCA}\label{sec:interaction}

The QCA defined in the previous section is gauge-invariant. However at this stage the addition of the gauge field has no dynamical effect on the electrons. This is best seen in the Heisenberg picture, where the the ``dressed'' creation operators $\overline a_x^\dagger$ and $\overline b_x^\dagger$ generate the same algebra, and evolve according to the same automorphism as the free Dirac QCA. Hence its continuum limit is also identical to that of the free Dirac QFT, but in terms of dressed creation operators, which, in turn, is identical to the Schwinger model defined by Eq.~(17) of Ref~\cite{melnikov2000lattice}, with zero charge: $g=0$.

This equivalence with the free Dirac QFT holds also for non-zero mass. Ref.~\cite{ArrighiDirac} rigorously proves that the continuum limit of the Dirac quantum walk, which is the one-particle sector of the QCA, yields the Dirac equation. The same is true for the full QCA \cite{Farrelly15}, and so we can consider that our dressed QCA is a reformulation of the Schwinger model defined by Eq.~(17) of Ref~\cite{melnikov2000lattice}, with zero charge.   

What is missing compared to the full Schwinger model is the local Hamiltonian term proportional to the square of the electric field: $E_x^2 = g^2 L_x^2$. It is this term that makes the model effectively interactive. 



\subsection{Schr\"odinger picture}\label{subsec:schrodinteraction}

Can we guess how to modify our QCA, so that the missing local Hamiltonian term shows up in continuum limit? Inspired by the many time-discretizations of Hamiltonian systems, one may be tempted to use the operator--splitting method and simply interleave an $\varepsilon$--period of evolution under the extra local Hamiltonian term, in-between any two time steps. For this to be justified by the Trotter-Kato formula, however, the free evolution over an $\varepsilon$--period ought to be approximately equal to the identity, too.

Let us examine whether this is the case. The gate $W'$ as defined is parameterized by the space/time lattice spacing $\varepsilon$ so as to yield free Dirac QFT in the limit $\varepsilon\rightarrow 0$. Observe that $\varepsilon$ only appears in the mass term, so that in the limit, $W'$ simply tends to its massless version, which is the swap. Hence the gate does not tend to the identity in the continuum limit, per se. 

Crucially, for the continuum limit to make sense, one must also make {\em states} dependent on the space-time spacing $\varepsilon$, in such a way that they are smooth at the scale of the lattice. In the single particle setting, this corresponds to bounds on the derivatives~\cite{ArrighiDirac}. For many-body states, one may require that they be approximately invariant under the permutation of nearby fermions. 

Hence, for a given value of $\varepsilon$, we may consider only those states $\ket{\psi}$ which are approximately invariant under the swap gate involved in the free Dirac dynamics, i.e. $W'\ket{\psi} \approx \ket{\psi}$ so that $W'\approx Id$ in the appropriate subspace.

Hence, we do propose to modify the QCA by simply interleaving a step resulting from the integration of the Hamiltonian $ \frac 1 2 \varepsilon \sum_x g^2 L_x^2$ (where $\varepsilon$ comes from the discretizing of the integral in Eq.~\eqref{SchwingerHam}), yielding the new gate
\begin{align}
\label{eq:full-qca}
&W''= \left(
\begin{array}{cccc}
{I} & 0 & 0 & 0\\
0 & -\ii s {I} & c {V} & 0\\
0 & c {V}^\dagger& -\ii s {I}& 0\\
0 & 0 & 0 & -{I}
\end{array}
\right)e^{\frac \ii 2 \varepsilon^2 g^2 {L}^2}
\end{align}
with ${L}\ket{l}=l\ket{l}$. 
This is clearly still gauge-invariant, since the operators $L_x$ are gauge-invariant observables. 

It is interesting to note that if ${L}^2=4\pi/g^2\varepsilon^2$ exactly, the phase wraps up around $2\pi$. Since the spectrum of $L$ is in $\mathbb Z$, this can only happen if $\varepsilon^2=(4\pi/g^2)/k$ with $k$ an integer. But, if we restrict ourselves to values of $\varepsilon$ such that this is the case, then we no longer need the whole Hilbert space ${\cal H}_{\mathbb{Z}}$ to represent the gauge field at each point : we can replace it by a $k$-dimensional Hilbert space instead. This, however, still requires that $k \longrightarrow \infty$ as $\varepsilon \longrightarrow 0$. This idea of restricting the gauge field to finite-dimensions labelling roots of unity is not new and has been evaluated in \cite{MagnificoFiniteEM}.


We write $\ket 0$ for the vacuum of the modes $a_x$ and $b_x$, tensored with any fixed joint eigenstate of the field operators $L_x$, for all lattice sites $x \in \varepsilon \mathbb Z$.

It is easy to see that the states created by applying any number of time the ``dressed'' fermionic creation operators ${\overline a_x}^\dagger$ and ${\overline b_x}^\dagger$ to $\ket 0$, for any $x$, are eigenstates of $e^{\frac \ii 2 \varepsilon^2 g^2 {L}^2}$. Hence, when considering these states, our particles still move as free particles, dragging the associated electric field with them. However, they now take a phase whose dynamics depends on the energy stored in that electric field. 


To observe the effect of the interaction, we need to consider wavefunctions that are smooth with respect to the lattice spacing. For instance, consider an external electric field in the shape of a step function, i.e. let $\ket 0$ be such that $L_x \ket 0 = 0$ for all $x<0$, and $L_x \ket 0 = -\ket 0$ for $x \ge 0$, so that if we create a particle at any position $x \ll 0$ with $\overline a_x$, then the electric field is still zero far away on both sides of space. 

Let us consider a single particle wavepacket
\begin{equation}
\ket{\psi} := \sum_{x \in \varepsilon \mathbb Z} e^{i p x} f(x) \, \overline a_x^\dagger \, \ket 0,
\end{equation}
for some smoothly varying envelop $f$ satisfying $f(0) \simeq 0$. Then the expectation value of the momentum operator $-i\frac{\partial}{\partial x}$ is approximately equal to $p$. A typical choice of envelop is the gaussian amplitude
\begin{equation}
f(x)=\frac{1}{(2\pi\sigma^2)^{\frac 1 4}}\exp\Bigl(-\frac{(x-x_0)^2}{4\sigma^2}\Bigr)
\end{equation}
with $x_0\ll 0$ the mean and $\sigma>0$ the standard deviation {chosen so that $1/\sigma\ll p$ and yet $x_0 + \sigma \ll 0$}. For this choice of envelop and for an adequate superposition of chiralities, \cite{ST12} shows that in the non-relativistic limit $p \ll m$, the wavepacket's expected position will move under the free Dirac equation, with constant velocity $v = p/m$. Because $W'$ implements free Dirac QFT, this should still hold. But now $W''$ will precede each step by $e^{\frac \ii 2 \varepsilon^2 g^2 {L}^2}$, whose effect is to add a phase
\begin{equation}
\label{eq:elecaccel}
\begin{split}
\ket{\psi'} &= \sum_{x \in \varepsilon \mathbb Z} e^{i p x - \frac i 2 \varepsilon g^2 x} f(x) \, \overline a_x^\dagger \, \ket 0,\\
&=\sum_{x \in \varepsilon \mathbb Z} e^{i p'x} f(x) \, \overline a_x^\dagger \, \ket 0.
\end{split}
\end{equation}
Thus, $p'= p - \frac 1 2 \varepsilon g^2$ the new expected momentum after an $\varepsilon$--period of time. This means that the particle is accelerating, according to an external force equal to $g E$, where $E = -\frac 1 2 g$ is the effective electric field that it feels given our boundary conditions.

\subsection{Continuum limit}
\label{cont-lim}

In order to prove that we recover the Schwinger model, we need to take a continuum limit of our system. This is a hard problem for any interacting many-body system, as it usually requires one to be able to ``solve'' the theory. This difficulty also appears at the core of standard quantum field theories, where a finite scale (minimal length or maximal energy, called a regulator) usually has to be introduced in order to obtain finite predictions. Then the parameters of the theory have to be made dependent on this regulator in such a way that the predictions converge as the regulator tends to zero or infinity  (whichever corresponds to the continuum limit). This procedure is known as {\em renormalization}. But doing this exactly would require being able to fully integrate the theory, so as to find the dependence of the predictions on the regulator. 

In the standard QFT formalism, this is done by defining the theory perturbatively, starting from a solvable model (typically a gaussian, i.e., a quasi-free theory defined by a Hamiltonian that is quadratic in the canonical variables).  The continuum limit is then taken independently for each order in perturbations. 

This means that we first need to establish the continuum limit for a set of parameters where we expect to arrive at an exactly solvable QFT. Thankfully, the Schwinger model can be solved when the fermions are massless, provided that one works with the Hilbert space formed by creating a finite number of fermions over a special ``vacuum'' state, which is Dirac's half-filled electron sea. 

However, this ``solution'' takes the form of an entirely different theory: a theory of free bosons with mass equal to $\frac {g} {\sqrt \pi}$. Since that theory is solvable in the sense that we can compute any $n$-point functions, all we need to do is to figure out which operators in the algebra of our QCA are those effective bosonic operators, and to show that they behave that way. 

{One may think of the choice of vacuum as a type of boundary condition: since we consider the Hilbert space spanned by acting on the vacuum only with $\mathfrak A$ the quasi-local algebra generated by Eqs. \eqref{eq:localg1}--\eqref{eq:localg5} (i.e. the balanced even polynomials of $\overline a_x$, $\overline b_x$ and $L_x$), it follows that of our states look far away just like the vacuum does.}  Moreover, the bosonic commutation relations and Klein-Gordon dynamics are true only weakly with respect to this Hilbert space, i.e., the equations hold only in terms of expectation values with respect to these states.

We will not attempt to given a rigorous proof of the continuum limit here. We just sketch the argument. Our approach is inspired by the derivation of the solution in Ref.~\cite{melnikov2000lattice}.

Let us therefore consider the massless case: $m=0$, and the QCA with $W''$ as defined in Eq.~\eqref{eq:full-qca}. At zero coupling (charge) $g=0$, The Heisenberg equations of motions are {\em exactly} those of the free Dirac QFT, except for the discrete restrictions of the space and time variables. In the continuum, however, the Hilbert space is usually restricted to the Fock space of the following annihilation operators:
\begin{equation}
c_p = \begin{cases}
      a_p, & \text{if}\ p \ge 0 \\
      b_p, & \text{if}\ p < 0 \\
    \end{cases}
    \quad \text{and}\quad 
    d_p = \begin{cases}
      b_p^\dagger, & \text{if}\ p \ge 0 \\
      a_p^\dagger, & \text{if}\ p < 0 \\
    \end{cases}
\end{equation}
where 
\begin{align}
a_p &:= \varepsilon \sum_x \frac 1 {\sqrt \varepsilon}\, \ol a_x e^{i p x}\\
b_p &:= \varepsilon \sum_x \frac{1}{\sqrt \varepsilon}\, \ol b_x  e^{i p x},
\end{align}
and $p$ takes value on the circle $T_\varepsilon$ of circumference $2\pi/\varepsilon$. 
So that, for instance,
\begin{equation}
a_p^\dagger a_q + a_q a_p^\dagger = 2 \pi \delta(p-q) I
\end{equation}
where 
\begin{equation}
\delta(p) = \frac{\varepsilon}{2\pi}\sum_x e^{ixp}
\end{equation}
is the Dirac delta over the circle:
\begin{equation}
\int_{T_\varepsilon} dp \, \delta(p) f_p = \int_{-\pi/\varepsilon}^{\pi/\varepsilon} dp\, \delta(p) f_p = f_0.
\end{equation}
The inverse Fourier transform is
\begin{equation}
\frac 1 {\sqrt \varepsilon} \,\ol a_x = \int_{T_\varepsilon}  \frac{dp}{2\pi} \, e^{-ipx} a_p.
\end{equation}

Also, the Kronecker delta $\delta_{x}$ defined for $x \in \varepsilon \mathbb Z$ is
\begin{equation}
\delta_x = \varepsilon  \int_{T_\varepsilon} \frac{dk}{2\pi} \, e^{ik(x-y)}.
\end{equation}


It is easy to check that one step of the QCA defined by $W'$ (or $W''$ with $g=0$) yields the Heisenberg dynamics
\begin{equation}
c_p \mapsto e^{-i \varepsilon |p| } c_p \quad\text{\and}\quad  d_p \mapsto e^{-i \varepsilon |p| } d_p.
\end{equation}

Also, we have
\begin{equation}
[\alpha \ol a_x + \beta \ol b_x, L_y] = \begin{cases}
      \alpha \ol a_x + \beta \ol b_x, & \text{if}\ y > x \\
      0, & \text{else} \\
    \end{cases}
\end{equation}
for any $\alpha, \beta \in \mathbb C$.

Let us consider the density of the left-moving fermion modes:
\begin{equation}
\begin{split}
A_x 
&= \frac 1 \varepsilon a_x^\dagger a_x 
=  \frac 1 \varepsilon \ol a_x^\dagger \ol a_x 
\end{split}
\end{equation}
Clearly, $[A_x, L_y] = 0$, so that the interaction ($g \neq 0$) has no bearing on its evolution at all.
Instead, let us consider a smeared version of it, by cutting off hight momentum modes. 

First, consider its Fourier transform
\begin{equation}
\begin{split}
A_k &= \varepsilon \sum_x e^{ikx} A_x \\
&= \int_{T_\varepsilon \times T_\varepsilon} \frac{dp\, dq}{2\pi} \, \delta(p+k-q)  a_p^\dagger a_{q} \\ 
&= \int_{T_\varepsilon} \frac{dp}{2\pi} \, a_p^\dagger a_{p+k} 
\end{split}
\end{equation}

We could define a smeared density such as
\begin{equation}
\label{eq:cgbad}
\tilde A_x^\Lambda = \int_{-\Lambda}^\Lambda \frac{dk}{2\pi} \, e^{-ikx} A_k 
\end{equation}
for some cutoff $\Lambda < \pi/\varepsilon$. But this clearly still commutes with $L_y$.

Instead, let us first smear the field operators themselves, and then use those to define the charge density.
For instance, we could use 
\begin{equation}
\psi^+_\Lambda(x) := \int_{-\Lambda}^\Lambda \frac{dp}{2\pi} \,a_p \, e^{-ipx},
\end{equation}
which is meant to tend to an actual Dirac field operator in the continuum. We then define the corresponding fermion density operator
\begin{equation}
A_x^\Lambda := (\psi^+_\Lambda(x))^\dagger \psi^+_\Lambda(x).
\end{equation}
Similarly, we define define $\psi^-_\Lambda(x)$ and $B_x^\Lambda$ from $\overline b_x$.

Let us expand $A_x^\Lambda$ in momentum to see how it differs from $\tilde A_x^\Lambda$. We have
\begin{equation}
A_x^\Lambda = \int_{T_\varepsilon} \frac{dk}{2 \pi}\, A_k^\Lambda,
\end{equation}
with Fourier transform
\begin{equation}
A_k^\Lambda = \int_{B_\Lambda} \frac{dp\,dq}{2\pi} \delta(p+k-q)\, a_p^\dagger a_q.
\end{equation}
where the double momentum integral is over the region $B_\Lambda = [-\Lambda,\Lambda]\times[-\Lambda, \Lambda]$.
Observe that this is just $A_x$ if $B_\Lambda = T_{\varepsilon}$. 
More generally, the region $B_\Lambda$ could be of any shape provided that it is bounded by a ball of radius that is a fixed multiple of $\Lambda$, and is symmetrical with respect an exchange of $p$ and $q$, so as to yield a self-adjoint operator. Proper convergence may also require replacing the sharp region by a smearing function with coefficient decaying rapidly beyond $\Lambda$, such as a Gaussian.


Let us consider the coarse-grained density of both types of fermions:
\begin{equation}
\label{eq:cggood}
\rho_q^\Lambda = A_q^\Lambda + B_q^\Lambda.
\end{equation}
We will also refer to the non-smeared density
\begin{equation}
\rho_q = A_q + B_q.
\end{equation}
If $f$ denotes the Heisenberg automorphism for the QCA with no coupling ($g=0$), then
\begin{align}
f(a_p^\dagger a_q) &=  e^{-i \varepsilon (q-p)} a_p^\dagger a_{q} ,\\
f(b_p^\dagger b_q) &=  e^{i \varepsilon (q-p)} b_p^\dagger b_{q}.
\end{align}
Hence
\begin{align}
f(A_k^\Lambda) &=  e^{-i \varepsilon k} A_k^\Lambda ,\\
f(B_k^\Lambda) &=  e^{i \varepsilon k} B_k^\Lambda.
\end{align}
Using the finite difference
\begin{equation}
\Delta^2_t \, \rho_q^\Lambda := \frac 1 {\varepsilon^2}( f^2(\rho_q^\Lambda) + \rho_q^\Lambda - 2 f(\rho_q^\Lambda) ), 
\end{equation}
we obtain
\begin{equation}
\Delta_t^2 \rho_q^\Lambda = - q^2 \rho_q^\Lambda + \mathcal O(\varepsilon).
\end{equation}
To lowest order in $\varepsilon$, this is the massless Klein-Gordon equation. 

The error term still has to be appropriately bounded to show convergence, but we will leave this for further work. Instead, let us consider how this equation changes in the presence of an interaction $g \neq 0$. If our system behaves like the Schwinger model, we should see an effective mass term appear in the Klein-Gordon equation.

Let $L_q = \varepsilon \sum_{x} L_{x+\frac 1 2 \varepsilon} e^{ipx}$ (where the sum is over $x \in \varepsilon \mathbb Z$), then,
\begin{equation}
\begin{split}
[\nol a_p,L_q] 
&= \varepsilon^{3/2} \sum_{x} \sum_{y \ge x}  e^{iqy} e^{i p x} \nol a_x\\
&=  \varepsilon^{3/2} \sum_{z\ge 0} e^{iqz} \sum_x e^{i (p+q) x} \nol a_x\\
&= \theta_q \,\nol a_{p+q},
\end{split}
\end{equation}
with $\theta_q := \varepsilon \sum_{y\ge 0} e^{iqy}$.
It follows that
\begin{equation}
[\nol a_p, L_q^\dagger L_q] = |\theta_q|^2\,\nol a_p + \theta_{-q} \, L_q \nol a_{p-q} + \theta_q \, L_{-q} \nol a_{p+q},
\end{equation}
where we used that fact that $L_q^\dagger = L_{-q}$ since $L_x$ is self-adjoint. 
We obtain the same equation substituting $\nol b_p$ for $\nol a_p$.

The electric field Hamiltonian is
\begin{equation}
H_E := \varepsilon \tfrac 1 2 g^2 \sum_{x} L_x^2 = g^2 \int_{T_\varepsilon} \frac{dk}{2\pi}\, L_k^\dagger L_k.
\end{equation}
Then
\begin{equation}
\begin{split}
[a_p^\dagger a_q, H_E] &= g^2 \int_{T_\varepsilon} \frac{dk}{2\pi} \, \bigl( \theta_k a_p^\dagger L_{-k} a_{q+k} - \theta_{-k} a_{p+k}^\dagger L_{k} a_q \bigr).
\end{split}
\end{equation}

Let us consider the evolution of the coarse-grained electron density given by Eq.~\eqref{eq:cggood}. We have
\begin{equation}
\label{eq:almostthere}
\begin{split}
[A_l^\Lambda,H_E] &= g^2 \int_{-\pi/\varepsilon}^{\pi/\varepsilon} \frac{dk}{2\pi} \,\theta_k \int_{B_\Lambda} \frac{dp dq}{2\pi}\,\delta(p+l-q) \\
&\quad \quad \times \bigl( a_p^\dagger L_{-k} a_{q+k} - a_{p-k}^\dagger L_{-k} a_q \bigr) \\
\end{split}
\end{equation}
(We applied the transformation $k \rightarrow -k$ on the second term, which does not change the value of the expression).
If there is no restriction on the integral in $p$ and $q$, this is zero because the first and second term differ only by a symmetry $p \rightarrow p-k$ of the integration domain. However, this argument fails thanks to the limited domain $B_\Lambda$.

We can simplify this expression by carefully choosing the dependence of $\Lambda$ on $\varepsilon$, and constraining the states to be sufficiently "smooth". For instance, suppose we consider states in the Fock space of $\ol a_x$, $\ol b_x$ for all $x$. Moreover, let us consider a state $\ket \psi$ in which only modes $p$ with $|p| < \delta \Lambda$ are occupied, where $0 < \delta < 1$. Then the expectation value of the integrand is zero for all $|k| < (1-\delta) \Lambda$. 
The terms in the integral with larger $k$ can be made to vanish in expectation if we also impose that the states of the electric field are smooth in that they contain no mode with $k$ above this bound. Consequently, the whole commutator is again zero in expectation value. 

That is, as we tend to the continuum by decreasing $\varepsilon$ and increasing $\Lambda$ proportionally to $\pi/\varepsilon$, we also consider fermion wavefunctions and electric field wavefunctions which are progressively smoother. (We leave open for now the question of whether this constraint is preserved by the dynamics).

To have an example where the commutator simplifies without entirely vanishing in this continuum limit, let us instead do this for the Fock space of the electron and positron annihilation operators $c_p$ and $d_p$, as mentioned in the beginning of this section.

To see which terms of the commutator survive, we need to commute operators such that either $c_p$ or $d_p$ are on the right side of monomials (or their adjoint on the left-side), so that they annihilate the state $\ket \psi$ if the modes $p$ are not occupied.
In the derivation below, we assume that, as above, the cutoffs $\Lambda$, $\varepsilon$ and $\ket \psi$ are such that this allows such term to be translate by $k$ without affecting their expectation values once integrated.



Let us focus on the first term, and abbreviate the integral over $p$ as a sum for a more compact notation. Below, the symbol $\approx$ indicates that we removed a term using the fact that it can be translated by $k$ and cancelled with the corresponding term stemming from the equivalent manipulations on $a_{p-k}^\dagger L_{-k} a_q$ in the expression above.

We have, where all integrals are over $dp$,
\begin{align*}
\int a_p^\dagger & L_{-k} a_{p+l+k} = \bs \int\displaylimits_{p+l+k<0} \bs a_p^\dagger L_{-k}  d^\dagger_{p+l+k} + \bs\int\displaylimits_{p+l+k \ge 0} \bs a_p^\dagger L_{-k}  c_{p+l+k} \\
&\approx \bs \int\displaylimits_{p+l+k<0}\bs  a_p^\dagger L_{-k} d^\dagger_{p+l+k}\\
&= \bs \int\displaylimits_{\substack{p+l+k<0\\ p<0}} \bs d_p L_{-k} d^\dagger_{p+l+k} + \bs \int\displaylimits_{\substack{p+l+k<0\\ p \ge 0}} \bs c_p^\dagger L_{-k} d^\dagger_{p+l+k}\\
&\approx \bs \int\displaylimits_{\substack{p+l+k<0\\ p<0}} \bs d_p L_{-k} d^\dagger_{p+l+k} \tag{\stepcounter{equation}\theequation}\\
&= \bs \int\displaylimits_{\substack{p+l+k<0\\ p<0}} \bs \Bigl( L_{-k} d_p d^\dagger_{p+l+k} - \theta_{k} a_{p+k}^\dagger d^\dagger_{p+l+k}  \Bigr)  \\
&= \bs \int\displaylimits_{\substack{p+l+k<0\\ p<0}} \bs \Bigl( - L_{-k} d^\dagger_{p+l+k} d_p - \theta_{k} a_{p+k}^\dagger d^\dagger_{p+l+k}  \Bigr) \\
\end{align*}
If $k + l \neq 0$, we can just anti-commute $d^\dagger_{p+l+k}$ and $d_p$, to get
\begin{equation}
\begin{split}
\int\displaylimits & a_p^\dagger L_{-k} a_{p+l+k} 
\;\approx\; \bs \int\displaylimits_{\substack{p+l+k<0\\p<0}} \bs \theta_{k} a_{p+k}^\dagger d^\dagger_{p+l+k} \\
&= \bs\int\displaylimits_{\substack{p+l+k<0\\ p<0, \,p+k<0}} \bs \theta_{k} d_{p+k}d^\dagger_{p+l+k}
+ \bs \int\displaylimits_{\substack{p+l+k<0\\ p<0, \,p+k\ge 0}} \bs \theta_{k} c_{p+k}^\dagger d^\dagger_{p+l+k} \\
& \approx \bs \int\displaylimits_{\substack{p+l+k<0\\ p<0, \,p+k<0}} \bs \theta_{k} d_{p+k}d^\dagger_{p+l+k} \\
&= \bs \int\displaylimits_{\substack{p+l+k<0\\ p<0, \,p+k<0}} \bs - \theta_{k}d^\dagger_{p+l+k}  d_{p+k}  \;\;\approx\;\; 0,
\end{split}
\end{equation}
where we also assumed $l \neq 0$.

It is when $k+l = 0$ that a non-trivial term pops up, namely
\begin{equation}
\begin{split}
[A_l^\Lambda,H_E] &\;\;\simeq\;\; g^2 \int_{-\pi/\varepsilon}^{\pi/\varepsilon} \frac{dk}{2\pi} \,\theta_k \int_{-1/\Lambda}^0 dp\, \delta(k+l) L_{-k}\\
&\quad \quad - g^2 \int_{-\pi/\varepsilon}^{\pi/\varepsilon} \frac{dk}{2\pi} \,\theta_k \int_{-1/\Lambda}^k dp\, \delta(k+l) L_{-k},\\
&= g^2 \int_{-\pi/\varepsilon}^{\pi/\varepsilon} \frac{dk}{2\pi} \,\theta_k \,k\, \delta(k+l) L_{-k}
\end{split}
\end{equation}



Hence, with respect the Fock space with cutoff $\Lambda$, we obtain
\begin{equation}
[A_q^\Lambda, H_E] \;\;\simeq\;\; \frac{g^2}{2\pi} q\, \theta_{-q} L_{q}
\end{equation}
for $q \neq 0$. 
Since $[L_q,H_E] = 0$, then, given $U =  e^{i \varepsilon H_E}$,
\begin{equation}
U^\dagger A_q^\Lambda U \;\;\simeq\;\; A_q^\Lambda + i \varepsilon \frac{g^2}{2\pi}\, q\, \theta_{-q} L_q.
\end{equation}
Similarly
\begin{equation}
[B_q^\Lambda, H_E] \;\;\simeq\;\; - \frac{g^2}{2\pi} q\, \theta_{-q} L_{q}.
\end{equation}

We see that, since $\rho_q^\Lambda = A_q^\Lambda + B_q^\Lambda$,
\begin{equation}
[\rho_q^\Lambda, H_E] \;\;\simeq\;\; 0.
\end{equation}
Which is what we expect from the Schwinger model.

In the Heisenberg picture, one step of the QCA defined by $W''$ is $f_g(A) = f(U^\dagger A U)$ with $U$ defined as above, and $f$ is the QCA step defined by $W'$. 

One step of the full QCA is 
\begin{equation}
f_g(\rho_q^\Lambda) \;\;\simeq\;\; e^{- i \varepsilon q} A_q^\Lambda + e^{i \varepsilon q} B_q^\Lambda = f(\rho_q^\Lambda)
\end{equation}
We can compute a second step easily using the above results, and we obtain
\begin{equation}
\begin{split}
f^2_g(\rho_q^\Lambda) &\simeq  e^{-i 2 \varepsilon q} A_q^\Lambda + e^{i 2 \varepsilon q} B_q^\Lambda  \\
& \quad\quad + i\varepsilon \frac{g^2}{2\pi} q \,\theta_{-q} f(L_{q}) (e^{- i \varepsilon q} - e^{i \varepsilon q}). \\
\end{split}
\end{equation} 

We observe that
\begin{equation}
\begin{split}
 (e^{- i \varepsilon q} - e^{i \varepsilon q})\, \theta_{-q} &= \varepsilon \sum_{y \ge 0} (e^{i q (y-\varepsilon)} - e^{i q(y+\varepsilon)})\\
 &\simeq \varepsilon (1 + e^{-i q \varepsilon}) = 2 \varepsilon + \mathcal O(\varepsilon^2),
\end{split}
\end{equation}
provided that it multiplies operators which vanish at $x \rightarrow \infty$ when evaluated on states of our Fock space.

Also,
\begin{equation}
\begin{split}
f(L_{q}) 
&= \varepsilon \sum_x e^{i q x} (L_{x- \frac 1 2 \varepsilon} + \ol a_{x}^\dagger \ol a_x - \ol b_{x-\varepsilon}^\dagger \ol b_{x-\varepsilon})\\
&= L_q + \varepsilon (A_q - B_q). 
\end{split}
\end{equation}

But all states in our Fock space are gauge-invariant, i.e., $J_x \ket \psi = 0$. 
Hence,
\begin{equation}
\begin{split}
L_q &= \varepsilon \sum_x e^{i q x} L_{x+\frac 1 2 \varepsilon} \\
&= \varepsilon \sum_x e^{i q x} (L_{x-\frac 1 2 \varepsilon} + A_x + B_x)\\
&= e^{i q \varepsilon} L_q + \varepsilon( A_q + B_q ).
\end{split}
\end{equation}
Therefore,
\begin{equation}
-i q L_q = A_q + B_q + \mathcal O(\varepsilon).
\end{equation}
Finally,
\begin{equation}
f^2_g(\rho_q^\Lambda) \;\;\simeq\;\; f^2(\rho_q^\Lambda) - \frac{g^2} \pi (A_q + B_q),
\end{equation}
to lowest order in $\varepsilon$. Moreover, in expectation with respect to our ``smooth'' states, we expect that
\begin{equation}
A_q + B_q = \rho_p \;\simeq\; \rho_p^\Lambda.
\end{equation}

Since also $f_g(\rho_q^\Lambda) = f(\rho_q^\Lambda)$, then $\rho_q^\Lambda$ approximately satisfies the different equation
\begin{equation}
\label{eq:KG}
\Delta^2_t\, \rho_p^\Lambda \;\;\simeq\;\; - p^2 \rho_p^\Lambda - \frac{g^2} \pi \rho_p^\Lambda,
\end{equation}
which is the Klein Gordon equation with mass $g/\sqrt \pi$, as expected for the (massless) Schwinger model.

{Clearly, this derivation remains informal, but has the virtue of indicating the assumptions that may be required to obtain a rigorous continuum limit. Another step that we will not take here, would be to check that perturbation of this solution for a small mass also carries over to the continuum limit.}



\section{Conclusion}\label{sec:conclusion}

{\em Summary of achievements.} Fig.\ \ref{fig:QEDQCA} describes a quantum circuit made by infinitely repeating the local quantum gate given by \eqref{eq:full-qca}. This is a quantum cellular automaton (QCA), or rather a family of QCA, since the lattice spacing and local quantum gates are parameterized by $\varepsilon$. We have argued, in two ways, that this family of QCA constitutes an alternative formulation of Schwinger model, i.e. $(1+1)$--dimensional quantum electrodynamics (QED). We first did this by taking the limit when $\varepsilon\rightarrow 0$, and studying perturbations over suitable choices of `smooth' vacuum states, which allowed us to recover the main features of $(1+1)$--QED. Namely, electron acceleration under an electric field in \eqref{eq:elecaccel}, and the emergence of free bosons of effective mass $g/{\sqrt \pi}$ in \eqref{eq:KG} over Dirac's half-filled electron sea. The second argument is based on the fact that the very construction of the QCA was justified on the same grounds that the construction of $(1+1)$--QED is justified on. I.e.\ we started from an established QCA for the free theory (Fig.\ \ref{fig:DiracQCA} with \eqref{eq:DiracGate}), restored $U(1)$--gauge invariance by introducing the gauge field (Fig.\ \ref{fig:QEDQCA} with \eqref{eq:IntermGate}, which satisfies \eqref{eq:gaugeinv}), and eventually gave this gauge field a simple gauge-invariant dynamics.  

{\em QCA formulation of QFT.} The path-integral formulation of QFT suffers from a number of issues related to the quantization of the action, a process which breaks continuity, introduces ambiguities, and jeopardizes unitarity. We hope that the $(1+1)$--QED QCA hereby presented will serve as an illustration that QCA formulations of QFT are feasible and advantageous in these respects. Taking a natively quantum and discrete dynamics as the starting point directly shortcuts all of these issues. Getting back to the continuum is no extra work:\ renormalization had to be worked out in the path-integral formulation anyway. We showed that gauge-invariance could be handled straightforwardly, at least within the temporal gauge. In principle, Lorentz-covariance could also be verified using the approach form Ref.~\cite{ArrighiLORENTZ,PaviaLORENTZ,PaviaLORENTZ2, DebbaschLORENTZ}. Altogether the construction is therefore simple and pedagogical. The result is more in line with the discrete spacetime formalisms introduced in Quantum Gravity proposals \cite{RovelliLQG,LollCDT}. 

{\em Quantum simulation. } Again, this $(1+1)$--QED QCA is but a quantum circuit, see Fig.\ \ref{fig:QEDQCA}. Each $W''$ can be expressed in terms of standard universal gates such as $\textsc{CNot, Hadamard, Phase}$. Thus, the QCA is directly interpretable as a digital quantum simulation algorithm, to run on a Quantum Computer. Moreover, this quantum simulation algorithm is efficient, in the sense that it requires an $O(st/\varepsilon^2)$ gates in order to simulate a chunk of space of size $s$, over $t$ time steps, with $\varepsilon$ the spacetime resolution---which for free fermions is known to control the precision in $||\cdot||_2$--norm linearly \cite{ArrighiDirac}. Of course the output produced is a quantum state, and so one may then need to repeat the process several times in order to obtain useful statistics about it. 
Classically however, just the state space itself is already of size an $O(\exp(s/\varepsilon))$, as it grows exponentially with the number of quantum systems to be simulated---the classical time complexity is therefore at least an $O(\exp(s/\varepsilon)t/\varepsilon)$. Clearly, the exponential gain here is due to the fact that that the $(1+1)$--QED QCA simulates multi-particle systems, just like in Hamiltonian-based multi-particle quantum simulation schemes. QW-based quantum simulation schemes, on the other hand, are by definition in the one-particle sector, and thus can only yield polynomial gains.  

{\em State preparation and measurement. } When simulating quantum field theories, the preparation of states, and the identification of observables which match those relevant to a specific experiment, are challenging matters. One central issue is the fact that the nature of the {\em vacuum} for the full theory (lowest energy state) is not known a priori. This question of the state preparation for quantum simulation was considered specifically for fermions in \cite{JLP14}. Although this work uses a continuous-time approach, the same strategy should also work for QCA. The idea is to start from a known ``non-interacting vacuum'', and simulate an evolution where an interaction parameter is slowly (adiabatically) turned on with time. This works only if there is a sufficiently large energy gap above the ground state, which needs be unique for every value of the parameter. Typically, this parameter would be the charge, but in the specific example at hand the fermionic mass seems a good candidate. Indeed, in that case an energy gap is suggested by the mass of the effective bosons, in turn controlled by the electric charge.

For measurements, the prescription to measure local charge densities via phase estimation (\cite{JLP14} section 4.5) should carry over to our framework. Electric field operators are also local and again simple to measure. 
 

{\em Perspectives.} Regarding the $(1+1)$--QED QCA hereby presented, we hope that more detailed analysis will follow that will anchor the connection with the Schwinger model on firmer ground. Whilst recovering the full path integral formulation theory from the QCA seems out of reach (due to a lack the mathematical techniques for doing so), a more thorough derivation of its phenomenology is at hand. We also leave open the interesting question of determining the vacuum for a QCA, since there is no Hamiltonian. Here we chose it based on knowledge of the target continuum theory. However, given only a QCA, how would one go about arguing for a specific choice of vacuum? Eventually, the work needs to be extended in the obvious directions:\ Lorentz-covariance, higher spatial dimensions, and Yang-Mills theories. Plenty of fascinating questions lie ahead. 

\vspace{0.4cm}

\subsection*{Acknowledgments}

CB was supported by the National Research Foundation of Korea (NRF-2018R1D1A1A02048436). PA would like to thank Pablo Arnault, Nathana\"el \'Eon and Giuseppe Di Molfetta for helpful discussions.  TCF would like to thank Tony Short and Tobias Osborne for useful discussions. 







\bibliographystyle{plain}	
\bibliography{biblio}

\begin{thebibliography}{10}

\bibitem{ahlbrecht2012molecular}
Andre Ahlbrecht, Andrea Alberti, Dieter Meschede, Volkher~B Scholz, Albert~H
  Werner, and Reinhard~F Werner.
\newblock Molecular binding in interacting quantum walks.
\newblock {\em New Journal of Physics}, 14(7):073050, 2012.

\bibitem{LollCDT}
Jan Ambj{\o}rn, Jerzy Jurkiewicz, and Renate Loll.
\newblock Emergence of a 4d world from causal quantum gravity.
\newblock {\em Physical review letters}, 93(13):131301, 2004.

\bibitem{DebbaschWaves}
Pablo Arnault and Fabrice Debbasch.
\newblock Quantum walks and gravitational waves.
\newblock {\em arXiv preprint arXiv:1609.00722}, 2016.

\bibitem{arnault2016quantum}
Pablo Arnault, Giuseppe Di~Molfetta, Marc Brachet, and Fabrice Debbasch.
\newblock Quantum walks and non-abelian discrete gauge theory.
\newblock {\em Physical Review A}, 94(1):012335, 2016.

\bibitem{ArrighiGRDirac3D}
P.~Arrighi and S.~Facchini.
\newblock {Quantum walking in curved spacetime: $(3+1)$-dimensions, and
  beyond}.
\newblock {\em Pre-print arXiv:1609.00305}, 2016.

\bibitem{ArrighiGRDirac}
P.~Arrighi, S.~Facchini, and M.~Forets.
\newblock {Quantum walking in curved spacetime}.
\newblock {\em Quantum Information Processing}, 15:3467--3486, 2016.

\bibitem{ArrighiDirac}
P.~Arrighi, M.~Forets, and V.~Nesme.
\newblock {The Dirac equation as a Quantum Walk: higher-dimensions,
  convergence}.
\newblock Pre-print arXiv:1307.3524, 2013.

\bibitem{ArrighiQGOL}
P.~Arrighi and J.~Grattage.
\newblock {A quantum Game of Life}.
\newblock In {\em Second Symposium on Cellular Automata "Journ{\'e}es Automates
  Cellulaires" (JAC 2010), Turku, December 2010. TUCS Lecture Notes 13, 31-42,
  (2010).}, 2010.

\bibitem{ArrighiPQCA}
P.~Arrighi and J.~Grattage.
\newblock {Partitioned Quantum Cellular Automata are Intrinsically Universal}.
\newblock {\em Natural Computing}, 11:13--22, 2012.

\bibitem{ArrighiUCAUSAL}
P.~Arrighi, V.~Nesme, and R.~Werner.
\newblock {Unitarity plus causality implies localizability}.
\newblock {\em J. of Computer and Systems Sciences}, 77:372--378, 2010.
\newblock QIP 2010 (long talk).

\bibitem{ArrighiLORENTZ}
P.~Arrighi and C.~Patricot.
\newblock {A note on the correspondence between qubit quantum operations and
  special relativity}.
\newblock {\em Journal of Physics A: Mathematical and General},
  36(20):L287--L296, 2003.

\bibitem{ArrighigaugeRCA}
Pablo Arrighi, Giuseppe Di~Molfetta, and Nathana{\"e}l Eon.
\newblock A gauge-invariant reversible cellular automaton.
\newblock In {\em International Workshop on Cellular Automata and Discrete
  Complex Systems}, pages 1--12. Springer, 2018.

\bibitem{arrighi2014discrete}
Pablo Arrighi, Stefano Facchini, and Marcelo Forets.
\newblock Discrete lorentz covariance for quantum walks and quantum cellular
  automata.
\newblock {\em New Journal of Physics}, 16(9):093007, 2014.

\bibitem{arrighi2013dirac}
Pablo Arrighi, Vincent Nesme, and Marcelo Forets.
\newblock The dirac equation as a quantum walk: higher dimensions,
  observational convergence.
\newblock {\em Journal of Physics A: Mathematical and Theoretical},
  47(46):465302, 2014.

\bibitem{Bialynicki-Birula}
I.~Bialynicki-Birula.
\newblock {Weyl, Dirac, and Maxwell equations on a lattice as unitary cellular
  automata}.
\newblock {\em Phys. Rev. D.}, 49(12):6920--6927, 1994.

\bibitem{PaviaLORENTZ}
A.~Bibeau-Delisle, A.~Bisio, G.~M. D'Ariano, P.~Perinotti, and A.~Tosini.
\newblock Doubly special relativity from quantum cellular automata.
\newblock {\em EPL (Europhysics Letters)}, 109(5):50003, 2015.

\bibitem{PaviaMolecular}
Alessandro Bisio, Giacomo~Mauro D'Ariano, Paolo Perinotti, and Alessandro
  Tosini.
\newblock Thirring quantum cellular automaton.
\newblock {\em Physical Review A}, 97(3):032132, 2018.

\bibitem{PaviaLORENTZ2}
Alessandro Bisio, Giacomo~Mauro D’Ariano, and Paolo Perinotti.
\newblock Quantum walks, weyl equation and the lorentz group.
\newblock {\em Foundations of Physics}, 47(8):1065--1076, 2017.

\bibitem{CGW18}
C.~Cedzich, T.~Geib, A.~H. Werner, and R.~F. Werner.
\newblock {Q}uantum walks in external gauge fields, 2018.
\newblock arXiv:1808.10850v1.

\bibitem{DebbaschLORENTZ}
Fabrice Debbasch.
\newblock Action principles for quantum automata and lorentz invariance of
  discrete time quantum walks.
\newblock {\em arXiv preprint arXiv:1806.02313}, 2018.

\bibitem{DGDT06}
T.~DeGrand and C.~DeTar.
\newblock {\em Lattice Methods for Quantum Chromodynamics}.
\newblock World Scientific, Singapore, 2006.

\bibitem{DdV87}
C.~Destri and H.~J. de~Vega.
\newblock Light cone lattice approach to fermionic theories in 2-d: the massive
  {Thirring} model.
\newblock {\em Nucl.Phys.}, B290:363, 1987.

\bibitem{MolfettaDebbasch2014Curved}
Giuseppe Di~Molfetta, Marc Brachet, and Fabrice Debbasch.
\newblock Quantum walks in artificial electric and gravitational fields.
\newblock {\em Physica A: Statistical Mechanics and its Applications},
  397:157--168, 2014.

\bibitem{di2014quantum}
Giuseppe Di~Molfetta, Marc Brachet, and Fabrice Debbasch.
\newblock Quantum walks in artificial electric and gravitational fields.
\newblock {\em Physica A: Statistical Mechanics and its Applications},
  397:157--168, 2014.

\bibitem{di2016quantum}
Giuseppe Di~Molfetta and Armando P{\'e}rez.
\newblock Quantum walks as simulators of neutrino oscillations in a vacuum and
  matter.
\newblock {\em New Journal of Physics}, 18(10):103038, 2016.

\bibitem{EisertSupersonic}
Jens Eisert and David Gross.
\newblock Supersonic quantum communication.
\newblock {\em Physical review letters}, 102(24):240501, 2009.

\bibitem{MagnificoFiniteEM}
Elisa Ercolessi, Paolo Facchi, Giuseppe Magnifico, Saverio Pascazio, and
  Francesco~V. Pepe.
\newblock Phase transitions in ${Z}_{n}$ gauge models: Towards quantum
  simulations of the schwinger-weyl qed.
\newblock {\em Phys. Rev. D}, 98:074503, Oct 2018.

\bibitem{Farrelly15}
T.~C. Farrelly.
\newblock {\em Insights from Quantum Information into Fundamental Physics}.
\newblock PhD thesis, University of Cambridge, 2015.
\newblock arXiv:1708.08897.

\bibitem{farrelly2014causal}
T.~C. Farrelly and A.~J. Short.
\newblock Causal fermions in discrete space-time.
\newblock {\em Physical Review A}, 89(1):012302, 2014.

\bibitem{Feynman_chessboard}
Hibbs Feynman.
\newblock {\em Quantum mechanics and path integrals}.
\newblock McGraw-Hill, 1965.
\newblock feynman relativistic chessboard.

\bibitem{FeynmanQC}
R.~P. Feynman.
\newblock {Simulating physics with computers}.
\newblock {\em International Journal of Theoretical Physics}, 21(6):467--488,
  1982.

\bibitem{FeynmanQCA}
R.~P. Feynman.
\newblock {Quantum mechanical computers}.
\newblock {\em Foundations of Physics (Historical Archive)}, 16(6):507--531,
  1986.

\bibitem{FML17}
F.~Fillion-Gourdeau, S.~MacLean, and R.~Laflamme.
\newblock Algorithm for the solution of the dirac equation on digital quantum
  computers.
\newblock {\em Phys. Rev. A}, 95:042343, 2017.

\bibitem{georgescu2014quantum}
IM~Georgescu, Sahel Ashhab, and Franco Nori.
\newblock Quantum simulation.
\newblock {\em Reviews of Modern Physics}, 86(1):153, 2014.

\bibitem{HastingsMonteCarlo}
W~Keith Hastings.
\newblock Monte carlo sampling methods using markov chains and their
  applications.
\newblock {\em Biometrika}, 57(1):97--109, 1970.

\bibitem{JLP14}
S.~P. Jordan, K.~S.~M. Lee, and J.~Preskill.
\newblock {Q}uantum {A}lgorithms for {F}ermionic {Q}uantum {F}ield {T}heories,
  2014.
\newblock arXiv:1404.7115v1.

\bibitem{PreskillQuantumSim}
Stephen~P Jordan, Keith~SM Lee, and John Preskill.
\newblock Quantum algorithms for quantum field theories.
\newblock {\em Science}, 336(6085):1130--1133, 2012.

\bibitem{Kaplan09}
D.~B. Kaplan.
\newblock {C}hiral symmetry and lattice fermions, 2009.
\newblock arXiv:0912.2560v2.

\bibitem{HamiltonianBasedSchwinger}
N.~Klco, E.~F. Dumitrescu, A.~J. McCaskey, T.~D. Morris, R.~C. Pooser, M.~Sanz,
  E.~Solano, P.~Lougovski, and M.~J. Savage.
\newblock Quantum-classical computation of schwinger model dynamics using
  quantum computers.
\newblock {\em Phys. Rev. A}, 98:032331, Sep 2018.

\bibitem{QuantumClassicalSim}
N~Klco, EF~Dumitrescu, AJ~McCaskey, TD~Morris, RC~Pooser, M~Sanz, E~Solano,
  P~Lougovski, and MJ~Savage.
\newblock Quantum-classical computation of schwinger model dynamics using
  quantum computers.
\newblock {\em Physical Review A}, 98(3):032331, 2018.

\bibitem{kogut1975hamiltonian}
John Kogut and Leonard Susskind.
\newblock Hamiltonian formulation of wilson's lattice gauge theories.
\newblock {\em Physical Review D}, 11(2):395, 1975.

\bibitem{InnsbruckLGT}
Esteban~A Martinez, Christine~A Muschik, Philipp Schindler, Daniel Nigg,
  Alexander Erhard, Markus Heyl, Philipp Hauke, Marcello Dalmonte, Thomas Monz,
  Peter Zoller, et~al.
\newblock Real-time dynamics of lattice gauge theories with a few-qubit quantum
  computer.
\newblock {\em Nature}, 534(7608):516--519, 2016.

\bibitem{melnikov2000lattice}
Kirill Melnikov and Marvin Weinstein.
\newblock Lattice schwinger model: Confinement, anomalies, chiral fermions, and
  all that.
\newblock {\em Physical Review D}, 62(9):094504, 2000.

\bibitem{meyer1996quantum}
David~A Meyer.
\newblock From quantum cellular automata to quantum lattice gases.
\newblock {\em Journal of Statistical Physics}, 85(5-6):551--574, 1996.

\bibitem{NN81}
H.~B. Nielsen and M.~Ninomiya.
\newblock A no-go theorem for regularizing chiral fermions.
\newblock {\em Physics Letters B}, 105(2–3):219 -- 223, 1981.

\bibitem{Osborne19}
T.~J. Osborne.
\newblock {C}ontinuum {L}imits of {Q}uantum {L}attice {S}ystems, 2019.
\newblock arXiv:1901.06124v1.

\bibitem{ST12}
Sang~Tae Park.
\newblock Propagation of a relativistic electron wave packet in the dirac
  equation.
\newblock {\em Physical Review A}, 86(6):062105, 2012.

\bibitem{quigg2013gauge}
Chris Quigg.
\newblock {\em Gauge theories of the strong, weak, and electromagnetic
  interactions}.
\newblock Princeton University Press, 2013.

\bibitem{rico2014tensor}
E~Rico, T~Pichler, M~Dalmonte, P~Zoller, and S~Montangero.
\newblock Tensor networks for lattice gauge theories and atomic quantum
  simulation.
\newblock {\em Physical Review Letters}, 112(20):201601, 2014.

\bibitem{RovelliLQG}
Carlo Rovelli.
\newblock Simple model for quantum general relativity from loop quantum
  gravity.
\newblock In {\em Journal of Physics: Conference Series}, volume 314, page
  012006. IOP Publishing, 2011.

\bibitem{SchumacherWerner}
B.~Schumacher and R.~Werner.
\newblock {Reversible quantum cellular automata.}
\newblock arXiv pre-print quant-ph/0405174, 2004.

\bibitem{silvi2014}
Pietro Silvi, Enrique Rico, Tommaso Calarco, and Simone Montangero.
\newblock Lattice gauge tensor networks.
\newblock {\em New Journal of Physics}, 16(10):103015, Oct 2014.

\bibitem{BenziSucci}
Sauro Succi and Roberto Benzi.
\newblock Lattice boltzmann equation for quantum mechanics.
\newblock {\em Physica D: Nonlinear Phenomena}, 69(3):327--332, 1993.

\bibitem{ErezCiracLGT}
Erez Zohar, J~Ignacio Cirac, and Benni Reznik.
\newblock Quantum simulations of lattice gauge theories using ultracold atoms
  in optical lattices.
\newblock {\em Reports on Progress in Physics}, 79(1):014401, 2015.

\end{thebibliography}

\end{document}